\documentclass[11pt]{article}
\pdfoutput=1
\usepackage{jheppub}
\usepackage{graphicx}
\usepackage{hyperref}
\usepackage{amsmath}
\usepackage{threeparttable}
\usepackage{gensymb}
\usepackage{amssymb}

\newcommand{\ml}{\mathcal{L}}
\newcommand{\p}{\partial}
\newcommand{\ca}{\mathcal{A}}

\author{Andreas Karch and}
\author{Mianqi Wang}
\title{Universal Behavior of Entanglement Entropies in Interface CFTs from General Holographic Spacetimes}

\affiliation{University of Texas, Austin, Physics Department, Austin, TX, 78712, USA}

\emailAdd{karcha@utexas.edu}
\emailAdd{mqwang@utexas.edu}

\abstract{In previous work universal behavior was conjectured for the behavior of the logarithmic terms in the entanglement entropy of intervals in 1+1 dimensional interface conformal field theories (ICFTs). These putative universal terms were exhibited both in free field theories as well as a large class of holographic models. In this work we demonstrate that this same behavior in fact is realized in any holographic ICFT, significantly strengthening the case for the conjecture.
}

\date{\today}
\begin{document}

\maketitle

\section{Introduction}
\label{intro}

Entanglement entropy (EE) is a key quantity when trying to understand quantum mechanical states in the language of quantum information. It plays a crucial role in studies of fields as diverse as quantum gravity, materials science, and quantum computing. It is, however, often difficult to calculate. One of the few universal results about EEs is the famous formula \cite{Holzhey:1994we,Vidal:2002rm,Calabrese:2004eu} for the EE of an interval of length $l$ in a 1+1 dimensional conformal field theory (CFT) with central charge $c$, $S_{EE} = c/3 \log{l/\epsilon}$, where $\epsilon$ is a UV cutoff length scale. This result is universal in that it is completely insensitive to the dynamical details of the CFT and is solely driven by the central charge; the EE is determined by the conformal anomaly. A similar universal result can be obtained in the case of a CFT with boundary (BCFT). In this case the EE for an interval containing the boundary is found, using standard BCFT techniques \cite{Calabrese:2009qy,Cardy:1986gw,Cardy:2004hm}, to be \begin{equation} S_{EE} = \frac{c}{6} \log{l/\epsilon} + \log g ,\label{bcftee} \end{equation} where $g$ is the boundary entropy of \cite{Affleck:1991tk}. While $g$ encodes details of the boundary conditions and hence is sensitive to the dynamics, the term depending logarithmically on the length $l$ of the interval is once more universal.

A more interesting structure emerges in the case of interface CFTs (ICFTs), where two conformal field theories meet at codimension 1 defect. Of particular interest in this paper will be the case of a $1+1D$ ICFT, that is a point-like defect between a left CFT$_L$ and right CFT$_R$. While the symmetry of an ICFT is identical to that of a BCFT, it allows novel observables by considering intervals that extended asymmetrically on the two sides of the interface. An ICFT can always be mapped to a BCFT via the folding trick, replacing the ICFT by a single product BCFT of CFT$_L$ $\times$ CFT$_R$. However, only symmetric intervals in the ICFT map to a standard EE calculation in the folded BCFT. The general asymmetric interval is a genuinely new observable and therefore is not governed by the universal formula \eqref{bcftee}.

To be concrete, we're interested in the entanglement entropy of intervals inside this $1+1D$ ICFT, in several different configurations. Using notations from \cite{Karch:2021qhd}, to calculate the EE in an ICFT there are several scenarios of interest:
\begin{itemize}
\item case 1: Trace out degrees of freedom outside an interval of total length $l$ perpendicularly crossing the defect. The length of the interval on each side of the defect is $l_L$ and $l_R$. One special case is particularly interesting, and will show up in the universal relation below. We call it case 1b: $l_L=0,l_R=l$, the one-sided interval.

\item case 2: Trace out the degrees of freedom on one side of the defect, and what is left is the entanglement entropy between the CFT$_L$ and CFT$_R$. The entanglement entropy is UV and IR divergent and so depends both on a UV regulator $\epsilon$ and an IR regulator $L$.

\end{itemize}

Both of these well defined field theory quantities. All the scenarios in case 1 are conventional entanglement entropies, where we define a reduced density matrix for a subregion of space and define the entanglement entropy as the standard van Neumann entropy of this reduced density matrix. The entanglement entropy is UV divergent, and the divergence arises from short distance degrees of freedom near the edge of the subregion. Case 2 is somewhat more unusual. In spirit, one separates the degrees of freedom into the left and right CFT and traces out over one side. The resulting quantity is UV {\it and}  IR divergent and one needs to introduce separate UV and IR cutoffs. A precise algorithmic definition of this entropy can be given using the replica trick as has been proposed in \cite{Sakai:2008tt,Peschel_2005}. One may wonder whether case 2 is just a rephrasing of the more conventional case 1 entropies, with the IR cutoff corresponding to the interval size. This turns out not to be the case. As surveyed in \cite{Calabrese:2009qy} and reviewed below, the two give genuinely different answers even for the divergent terms in the EE which usually are universal. Also their holographic definitions make it clear that they are very different quantities. Nevertheless, we will find some universal relations between them which seem to suggest that, roughly speaking, case 2 can be thought of as an infinitely large interval where the spatial boundary $l_R$ is pushed all the way to infinity and instead the sum is regulated by an IR cutoff. But this is just an intuitive intepretation of the relation between two intrinsically defined quantities.

For now, let us focus on the situation with equal central charges on both sides, $c_L=c_R=c$. We will discuss cases with unequal central charges in section 4. In the symmetric case 1, namely, $l_L=l_R=l/2$, the folding trick works well, and the EE reduces to that of the BCFT, which has the standard result $S=\frac{2c}{6}\log l/\epsilon+\log g$ \cite{Cardy:1986gw}. This expression is analogous to the free CFT case without boundaries, which was derived using the 'folding trick' \cite{Calabrese:2009qy}. When the interval is not symmetric, the folding trick doesn't help and one has to rely on different methods. For special CFTs with a holographic dual the EE of ICFTs can still be calculated by utilizing the Ryu-Takayanagi (RT) formula \cite{Ryu:2006bv}.

\par Previous studies of ICFTs have found that in cases 1b and 2 the structure of the EE is much more interesting than in a BCFT, in that this time not even the coefficient of the logarithmic term is universal \cite{Sakai:2008tt, Peschel_2005}. Namely, it was found that for case 2 \cite{Sakai:2008tt}, $S=\frac{\sigma_2}{2}\log L/\epsilon+\log g$, while in case 1b we have \cite{Peschel_2005} $S=\sigma_1\log l/\epsilon+\log g$. The coefficients $\sigma_1$ and $\sigma_2$ depend on the transmission coefficient \cite{Bachas:2001vj} and hence the dynamics. They exhibit different limiting behavior as one takes the transmission coefficient to 0 or 1 by dialing the coupling constants governing the properties of the interface, so they are definitely not equal to each other. There does, however, appear to be a universal relation between these coefficients as demonstrated in \cite{Karch:2021qhd}
\begin{equation}
    \sigma_1=\frac{\sigma_2}{2}+\frac{c}{6}.
    \label{relation}
\end{equation}

\begin{figure}[h]
    \centering
    \includegraphics[width=1\textwidth]{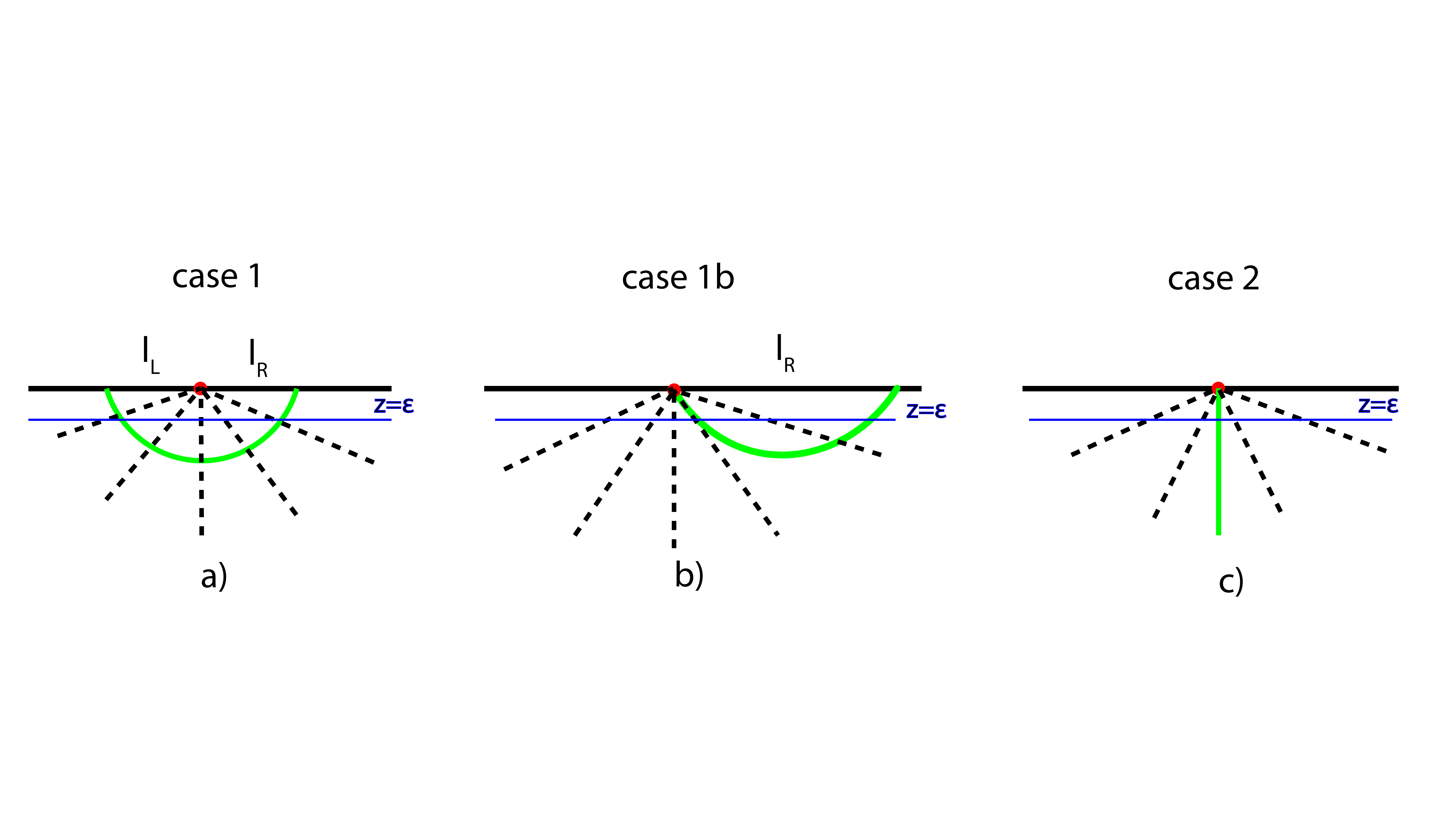}
    \caption{The configuration of intervals in the boundary ICFT and bulk RT surfaces for case 1, 1b and 2. The red dot is the interface, and green curves are RT surfaces. Black dash lines are constant $u_0$ slices. The blue solid lines are the cutoff $z=\epsilon$ in Poincare coordinates. a): generic case 1 with $l_L,l_R>0$. b): case 1b where $l_L=0$ and the RT surface is anchored right on the interface and the end of the interval. c): case 2 with RT surface extended to deep within the bulk, detecting EE of one side of the CFT. Notice the difference in values of $u_0$ for RT surfaces among those three scenarios: for generic case 1, $u_0$ goes through the entire range $(-\infty,\infty)$. For case 1b, it is bounded below, ranging $[u_{0C},\infty)$ in our notation (depends on internal coordinates when $d>0$). For case 2, it has a compact range, bounded on both sides. This is governed by the minimal surface ODE, and in general leads to different parametrizations for case 1 and case 2.}
    \label{fig:3rt}
\end{figure}

In \cite{Karch:2021qhd} this relation was found to be true in special holographic models, as well as in free field theories. Given that the relation held in both very strongly coupled as well as very weakly coupled examples, it was conjectured to be universally true. One should note that this universal relation only constrains the terms in the EE that scale logarithmically with the size $l$ of the interval (in case 1) and the IR cutoff (in case 2), or equivalently with the UV cutoff. No universal relation is expected for the subleading $l$ independent piece, the ICFT analog of $\log g$ in the BCFT. In the holographic setup, EE of the interval on the boundary ICFT is derived directly from the surface area of RT surfaces, which are codimensional-1 surfaces in space satisfying the minimal surface equation. Boundaries of RT surfaces are anchored at the boundaries of the interval. For case 1b, the RT surface is anchored at the interface defect on one end, and ends on the right boundary of the interval at position $l_R$ from the interface. For case 2, it is also anchored at the interface, but reaches out to infinity (with IR cutoff $L$), showing that all of the freedom in CFT$_2$ should be traced out. These two different RT surfaces are qualtitatively described in Figure \ref{fig:3rt}. Note that the holographic description makes it very clear that these are genuinely different quantities. In the case of a 3d holographic dual discussed in \cite{Karch:2021qhd}, the case 1 surface samples the entire warpfactor characterizing the 3d geometry, whereas the case 2 surface only is sensitive to the minimum of the warpfactor. What was argued there, and will be generalized here, is that nevertheless the divergent terms arising from the near boundary region in the two cases are related.

There is an intuitive picture that explains \eqref{relation}: one can argue that the contribution to the one-sided interval EE in $\sigma_1$ comes from two parts: one is the end point of the interval right on the interface, the other one is the other end point inside one of the CFTs. The former should give $\sigma_2/2$ since intuitively this is exactly half of the inter-CFT EE coefficient coming from the interface. The main part of this paper will be to establish this relation rigorously using the RT formulation. On the other side it is just the $c/6$ of ordinary EE coefficient of one boundary of the interval.

\par In \cite{Karch:2021qhd}, the relation \eqref{relation} was demonstrated analytically only in a special holographic setup, namely, an AdS$_2$ slicing of the asymptotic AdS$_3$ bulk dual to our $1+1D$ CFT. The method was to calculate the area of RT surfaces corresponding to the interval, through AdS$_2$ slicing inside the $2+1D$ spacetime. What was special in \cite{Karch:2021qhd} was that the size of the AdS$_2$ slices only depended on the radial coordinate of the asymptotic AdS$_3$ spacetime. To give a general proof of this relation in any ICFT with holographic dual, we should also consider the more general holographic bulk, which is an asymptotic AdS$_3$ with extra compactified dimensions, and allow dependence of the AdS$_2$ curvature radius also on the internal directions. We will start to see the effect of adding in the non-compact dimensions in section \ref{d1}, which will make the RT surfaces depend nontrivially in the compact dimensions. The special case of pure AdS$_3\times M^d$ is dual to the completely transparent interface, that is, a $1+1D$ CFT with no defect.

This more general holographic structure is in fact realized in many top-down string theory embeddings of ICFTs, as for example in \cite{Chiodaroli:2009yw} for the case of a holographic dual to a 1+1d ICFTs based on a type IIB compactification on AdS$_3$ $\times$ $S^3$ times an internal 4d Ricci flat manifold.

Concretely what we consider in this paper is a spacetime that is asymptotically AdS$_3$ correlated with a $d$-dimensional ($d\ge0$) smooth compact Riemannian manifold $M^d$ in space. We will calculate the minimal surface area of RT surfaces in this setup, which are codimension 1 surfaces in the $(2+1+d)D$ spacetime parametrized by the noncompact coordinate $u_0$ of the asymptotic AdS$_3$ and $d$ other compact coordinates $u_i,i=1,\dots,d$. This will give the coefficients of $\log l/\epsilon$ parts of EE of the intervals in case 1b) and case 2. Extending the results of \cite{Karch:2021qhd} to this most general holographic setting offers strong support for the universal relation (\ref{relation}).

\par For a $1+1D$ interface CFT with a conformal defect, we have the $(3+d)D$ holographic dual setup as mentioned above. The metric of the bulk spacetime is of the form
\begin{equation}
    ds^2=e^{2A(u_\mu)}\frac{dx^2-dt^2}{x^2}+g_{\mu\nu}du^\mu du^\nu.
    \label{metric}
\end{equation}

\begin{figure}[h]
    \centering
    \includegraphics[width=0.5\textwidth]{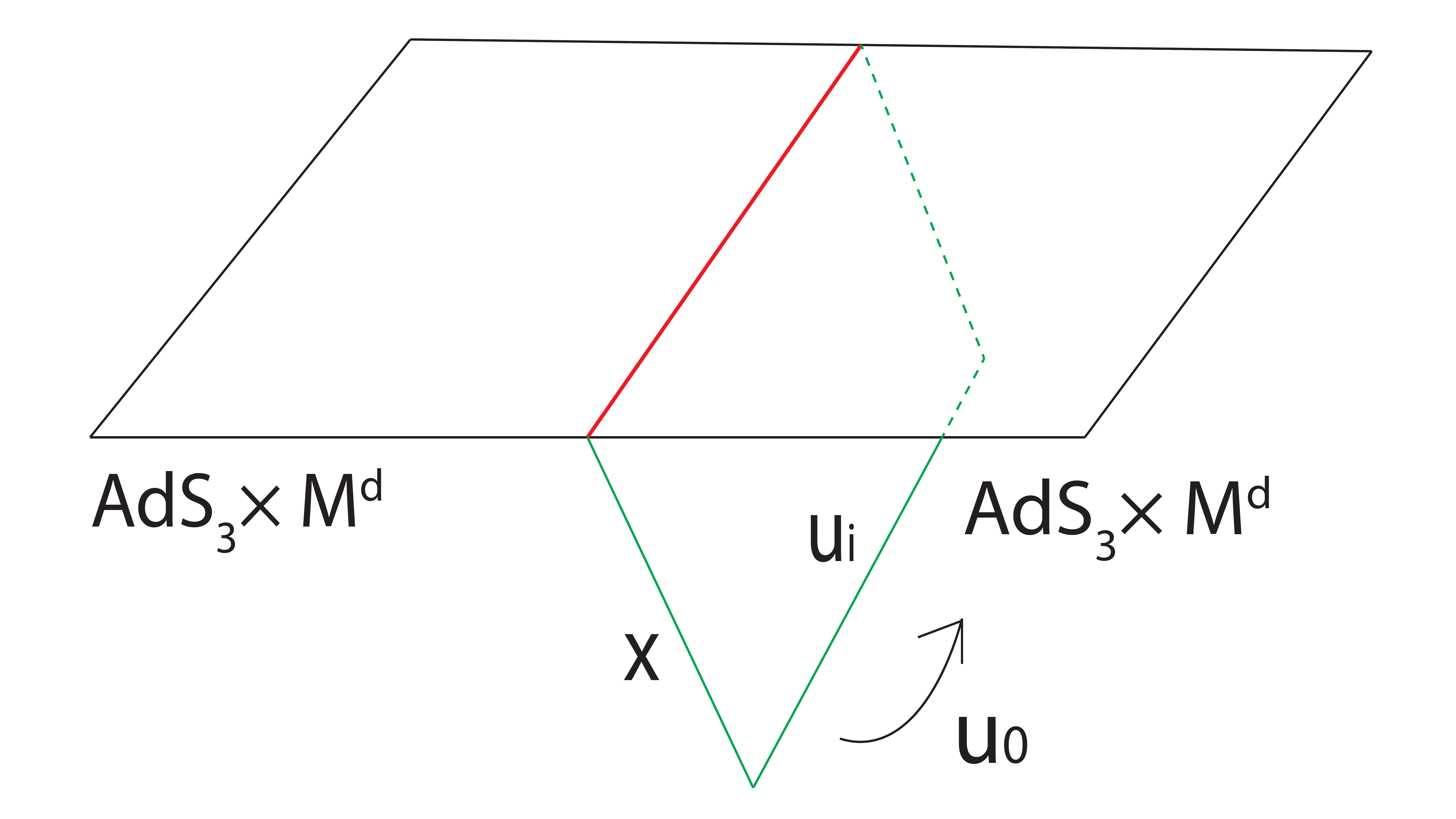}
    \caption{Spacetime in the bulk in our general holographic setup for $1+1$D ICFT is a warped product of AdS$_3$ and $M^d$. The compact internal manifold $M^d$ is sketched by the $u_i$ line, while the AdS$_3$ is sliced by AdS$_2$ with constant $u_0$ value (in green). Near the boundary, where the Poincare coordinate is at cutoff value $z=\epsilon$, the geometry splits to form the product AdS$_3\times M^d$. On the boundary, $M^d$ should be suppressed, leaving the $1+1$D ICFT. The red line in the middle (which should be a red dot) is the interface.}
    \label{fig:spacetime}
\end{figure}

The spacetime in our setup is an AdS$_2$ slicing (foliated by the non-compact $u_0$) of the asymptotic AdS$_3$ that emerges far from the defect line, coupled with a compact internal $C^\infty$-manifold $M^d$. More precisely the compact $M^d$ fibers over the asymptotic AdS$_3$. Throughout our paper, except for section \ref{uq}, the radius of the asymptotic AdS$_3$ will be $L=1$. The overall scale of each AdS$_2$ slice is characterized by the warpfactor $A(u_\mu)$, where $\mu,\nu=0,1,\dots,d$. For the internal Riemannian manifold $M^d$, the metric can be written as $g_{ij} du^{i} du^{j}$, where $i,j=1,\dots,d$. See Figure \ref{fig:spacetime} for a sketch of the spacetime. The warpfactor $e^{A(u_\mu)}$ satisfies: (1) $u_0$ runs from $-\infty$ to $\infty$; $A$ is a single-valued smooth function on $M^d$; (2) $e^A\sim \cosh(u_0-\delta u_\pm)$ as $u_0\rightarrow\pm\infty$ for some constants $\delta u_\pm$, and the spacetime there is the product form of a unit-radius  AdS$_3\times M^d$; (3) a minimal value of $A(u_0,u_i)$ at each fixed $u_i$ exists (denote as $A^*(u_i)$), and the critical point of $u_0$ as a function $u_{min}(u_i)$ is differentiable. We can set $u_{min}(0)=0$ as in the $d=0$ case.

As we will explain in details in the following sections, the general way to establish \eqref{relation} is as follows: First relate EE with the area $\ca$ of the corresponding $(d+1)$-dimensional RT surfaces in the bulk (henceforth we will only work on the $t=0$ slice). Then verify the contribution to (the leading logarithmically divergent terms of) $\ca$ in the two cases come from the near-boundary behavior of the RT surfaces, or the boundary $d$-dimensional curves on the compact internal manifold (up to some subtleties on the cutoff). We then identify those curves of case 2 and 1b near the interface, and calculate the contribution from those boundary curves of RT surfaces away from the interface to be $c/6$. Hence \eqref{relation} is proved.

We denote the near-interface boundary curves for RT surfaces in case 1b to be $u_{0C}(u_i)$. RT surfaces will be parametrized $x(u_\mu)$ in case 1, and the $d$-dimensional near-interface curves are obtained by taking the constant $x$ slice as $x\rightarrow 0$, or taking the zero locus of the denominator of the Lagrangian. The $d$-dimensional curve near the interface in case 2 is defined by $u_{0c}(u_i)\equiv u_0(x\rightarrow 0,u_i)$, since in case 2 RT surfaces will be parametrized as $u_0(x,u_i)$. The crucial step towards establishing \eqref{relation} will be proving the following key identity:
\begin{equation}
    u_{0c}(u_i)\equiv u_{0C}(u_i).
    \label{ki}
\end{equation}

\par Divergences will occur in our AdS$_3$/CFT$_2$ setup as $u_0$ goes to $\pm\infty$, so a UV cutoff must be introduced. We can transform the spacetime coordinates between AdS$_2$ and $2d$ Minkowski space on the slice. The same metric will then also naturally appear as the boundary metric the field theory lives on; from the boundary point of view these different coordinate systems in the bulk are related by a Weyl transformation of the boundary metric, which leaves the physics of a CFT invariant. This equivalence of 1+1d CFTs on AdS$_2$ versus 2d Minkowski space is however broken by the introduction of a cutoff. Cutoffs in AdS$_2$ will be position dependent when the cutoff in the $2d$ Minkowski is position independent, and vice versa. Here we are interested in a position independent cutoff in 2d Minkowski space, as this is the conventional choice in quantum field theory studies. Using the standard Poincare coordinates on AdS$_3$ and $e^A=\mathrm{cosh}(u_0)$, where the boundary is located at $z=0$, we have the Minkowski coordinates
\begin{equation}
    ds^2=\frac{1}{z^2}(-dt^2+dy^2+dz^2)
\end{equation}

where
\begin{equation}
    z=\frac{x}{\cosh u_0},\   y=-x\tanh u_0.
\end{equation}

Under assumption (2) above, as $u_0\rightarrow\pm\infty$ we know that the warpfactor goes to that of the AdS$_3$ case, so we will always have
\begin{equation}
    \frac{e^{u_0}}{2}=\frac{x}{z},
    \label{xz}
\end{equation}

so the position-independent UV cutoff in $2d$ Minkowski at $z=\epsilon$ is a position-dependent cutoff 
\begin{equation}
    u_c^+=\log\frac{2l_R}{\epsilon}
    \label{+co}
\end{equation}

in AdS$_2$, and on the other side 
\begin{equation}
    u_c^-=\log\frac{2l_L}{\epsilon}
    \label{-co}
\end{equation}

for $u_0\rightarrow -\infty$. Note that $l_L$ and $l_R$ are finite lengths indicating the interval in case 1, to be distinguished from the IR regulator $L$ appearing in the RT surface for case 2. The $\sigma_1$ and $\sigma_2$ we are about to determine are the coefficients of the $-\log \epsilon$ terms in the EE expressions in case 1b) and case 2, respectively.

\par The organization of this paper is as follows: We first do a quick review the $d=0$ case \cite{Karch:2021qhd} without compact dimensions in section \ref{d0}. In section \ref{d1} we demonstrate the case of AdS$_3 \times S^1$, or $d=1$. We derive the EOMs for RT surfaces in case 2 and case 1b respectively. In particular, we utilize different parametrizations on RT surfaces in the two cases. The key identity (\ref{ki}) will be carried out for the 1D curves $u_{0c}$ and $u_{0C}$ defined in the limit $x\rightarrow 0$ (that is near the interface). In section \ref{dd} the setup with general AdS$_3\times M^d$ bulk is analyzed, which is parallel to section \ref{d1}. We first demonstrate that we can always gauge fix diffeomorphism invariance by imposing a certain condition on the metric of the holographic bulk. Performing an analytic calculation of the EEs we find the universal relation to be true precisely under this condition. In section \ref{ex}, an example of a half-BPS six dimensional (0,4) supergravity solution will be treated analytically, and the universal relation associated with it be verified. In section \ref{uq}, we will discuss the case with unequal central charges on the left and right CFTs, and again verify our universal relation. Finally section \ref{ol} is the outlook to future research in this direction, in particular calculating the EE of general intervals in ICFT using the replica trick.  \\

\section{Review of $d=0$}
\label{d0}
First we do a quick review of the purely 3d asymptotically AdS$_3$ bulk ($d=0$). This part follows closely from \cite{Karch:2021qhd}. We rescale $u_0$ so that the only metric component $g_{00}=1$ in (\ref{metric}). We further assume that the warpfactor $e^A(u_0)$ reaches the minimal value $A^*$ at $u_0=0$. With no extra dimensions, RT surfaces are 1-dimensional curves, and their boundary 'curves' that we defined above are points (number values of $u_0$). In case 2, the RT surface starts off at $x=0$, and extends to $x\rightarrow \infty$, depicted by c) of Figure \ref{fig:3rt}. If for a given $x$ multiple values of $u_0$ occur, then the RT surface is not stable (i.e. does not satisfy the minimal area EOM). Another reason to use this parametrization is that in general, in the IR limit $x\rightarrow \infty$, $u_0$ value of the RT surface is bounded within a compact interval. In fact, we will see shortly below that if no extra dimensions are present, $u_0$ is a constant with no dependence on $x$. We can then parametrize it as $u_0(x)$. The Lagrangian (that is the area functional) of the RT surface is 
\begin{equation}
    \ml=\sqrt{(u_0')^2+\frac{e^{2A}}{x^2}}.
\end{equation}

The equation of motion (EOM) of this Lagrangian yields the extremal area condition for the RT surface. We can straightfowardly see that if $u_0'(x)=0$ and $A'(u_0)=0$, $\ml$ reaches its minima (or equivalently EOM for $\ml$ is satisfied), and that is indeed satisfied by $u_0(x)=0$, as predicted above. This corresponds to the RT surface fixed at a slice of AdS$_2$, or in our notation, $u_{0c}=0$. The entropy with UV cutoff $L$ and IR cutoff $\epsilon$ is then
\begin{equation}
    S=\frac{\ca}{4G_3}=\frac{e^{A^*}}{4G}\log\frac{L}{\epsilon}
\end{equation}

meaning that
\begin{equation}
    \sigma_2/2=\frac{e^{A^*}}{4G_3}=\frac{c}{6}e^{A^*},
\end{equation}

where $\ca$ is the area functional for the RT surface, and we have used the Brown-Henneaux (BH) relation \cite{Brown:1986nw}
\begin{equation} \label{brownhenneaux} G_3=3/2c. \end{equation} \\

\par In case 1, at constant time slice, the ending points for the RT surface are located at the boundary ICFT, see a), b) of Figure \ref{fig:3rt}. In generic case 1 with $l_L,l_R\neq 0$, the RT surface starts off at $u_0\rightarrow -\infty$ within CFT$_L$ and ends at $u_0\rightarrow \infty$ within CFT$_R$ (with proper UV cutoffs). In case 1b, the surface is anchored at the interface defect on one end, and at $x=l_R$ in CFT$_R$ on the other end. Taking the limit $x\rightarrow 0$ along the surface (or equivalently, taking the tangent vector of the RT surface in $(u_0,x)$ space at the interface $x=0$) gives us the lower bound of $u_0$ on the RT surface. Following our notation, this value is $u_{0C}$, and $u_0$ runs from $u_{0C}$ to $\infty$ in this case. Nevertheless, in either scenario we parametrize it using the well-defined function $x(u_0)$, since $u_0$ should be a monotonous variable along the RT surface (otherwise it seriously violate the minimal area condition). Alternatively, it is more matural to parametrize the RT surface using a smooth function whose codomain is compact, which is exactly the case with $x(u_0)$ since $\lim_{u_0\rightarrow\pm\infty}x(u_0)=l_{L/R}$. The similar parametrization holds for higher-dimensional generalization in the following sections. The Lagrangian for the RT surface is ($x'\equiv \partial_{u_0} x$)
\begin{equation}
    \ml=\sqrt{1+e^{2A}\frac{x'^2}{x^2}}.
\end{equation}

The scale isometry of AdS$_2$ corresponds to a symmetry $x\rightarrow \lambda x$. For an infinitesimal transformation $\lambda=1+\epsilon$, the Noether current $J_0$ for this symmetry is
\begin{equation}
    J_0(u_0)=-\frac{\delta\ml}{\delta x'}\frac{dx}{d\epsilon}=\frac{e^{2A}x'}{x\ml}.
\end{equation}

Since the conservation of charges is $\p_0J_0=0$, the current $J_0$ is an integration constant. Solving for $x'$ then gives the RT surface
\begin{equation}
    \frac{x'}{x}=\frac{J_0e^{-A}}{\sqrt{e^{2A}-J_0^2}},
    \label{es}
\end{equation}

and the on-shell Lagrangian reads
\begin{equation}
    \ml=\frac{1}{\sqrt{1-J_0^2e^{-2A}}}.
\end{equation}

We can see that for a given warpfactor $A(u_0)$ (and proper boundary conditions), the shape of the RT surface is exactly solvable. In particular for case 1b, as $u_0\rightarrow u_{0C}$ we should have $x\rightarrow 0$. From (\ref{es}) this tells us two things: 1) $u_{0C}=0$; 2) $J_0=e^{A^*}$. These are given by the facts that $J_0$ is constant, and that $A$ reaches its minimal value $A^*$ at $u_0=0$ by our aforementioned assumption. Indeed the $l_L\rightarrow 0$ limit is not smooth, since even an infinitesimally small $l_L$ will make $u_0$ run over the entire range.

In this simplest setup with no extra dimensions, (\ref{ki}) holds trivially $u_{0c}=u_{0C}=0$. According to the general dicussion in section \ref{intro}, it gives us \eqref{relation} as follows \cite{Karch:2021qhd}: We have already seen the expression for $\sigma_2/2$. The integration on $x$ to give the RT area functional $\ca$ (which in turn gives the log$\epsilon$ coefficient of EE) can be carried out trivially since $u_{0}= u_{0c}=0$ does no depend on $x$. This fact will remain true even for arbitrary extra dimensions. In particular, the cutoff near the interface is $x=e^{A^*}\epsilon$, and the coefficient for $-\log \epsilon$ is
\begin{equation}
    \sigma_2=\frac{c}{3}e^{A(u_{0c})}=\frac{c}{3}e^{A^*}.
\end{equation}

On the other hand, the method for determining $\ca$ in case 1 in general would be first doing a variation of $\ca$ with respect to $\delta l_L$ and $\delta l_R$

\begin{equation}
    \delta\ca=\ml|_{u_0=u_c^+}\frac{\delta u_c^+}{\delta l_R}\delta l_R-\ml|_{u_0=u_c^-}\frac{\delta u_c^-}{\delta l_L}\delta l_L.
    \label{var}
\end{equation}

Here $u_c^\pm$ are the (position dependent) UV cutoffs from (\ref{+co}), (\ref{-co}). Integrate with respect to $l_{L/R}$ and we find $\ca$
\begin{equation}
    \ca=\log l_L/\epsilon+\log l_R/\epsilon+const.
\end{equation}

Setting $l_L=\alpha l$ and $l_R=(1-\alpha) l$ with the total length of the interval $l=l_L+l_R$, we have the EE
\begin{equation}
    S=\frac{2}{4G_3}\log l/\epsilon +const. =\frac{c}{3}\log l/\epsilon +const.
\end{equation}

corresponding to a universal $c/6 \log l$ term on each end of the interval in the ICFT. For case 1b, the left end of the variation is at $u_{0C}^+$ instead of $u_c^-$. Here $u_{0C}^+$ is the cutoff, infinitesimally bigger than $u_{0C}$ (defined more carefully in the next section). Since $u_{0C}=0$, the left end variation will be again on the central AdS$_2$ leaf $u_0=0$, where the cutoff is similar to that in case 2, except with an $O\left((u_{0C}^+)^2\right)$ correction
\begin{equation}
    x=e^{A(u_{0C}^+)}\epsilon=(e^{A^*}+O\left((u_{0C}^+)^2\right))\epsilon.
\end{equation}

The Lagrangian at cutoff $u_{0C}^+$ is 
\begin{equation}
    \ml|_{u_0=u_{0C}^+}=\frac{e^{A^*}}{\sqrt{e^{2A(u_{0C}^+)}-e^{2A^*}}}+O\left(u_{0C}^+\right).
\end{equation}

The calculation of $\frac{\delta u_{0C}^+}{\delta l}$ involves a trick by integrating both sides of \eqref{es}. Integrate the left hand side as $(\log x)'$ from $x=e^{A(u_{0C}^+)}\epsilon$ to $x=l_R$, and the right hand side from $u_0=u_{0C}^+$ to $u_0=u_c^+$. Then take an $l$ derivative on both sides, we will get (see (2.38)-(2.40) of \cite{Karch:2021qhd})
\begin{equation}
    \frac{\delta u_{0C}^+}{\delta l}=-\frac{\sqrt{e^{2A(u_{0C}^+)}-e^{2A^*}}}{l}+O\left(u_{0C}^+\right).
\end{equation}

The variation of the area term is then
\begin{equation}
    \frac{\delta \ca}{\delta l}=\frac{1}{l}+\frac{e^{A^*}}{l},
\end{equation}

which makes the coefficient of $\log\epsilon$ for EE
\begin{equation}
    \sigma_1=\frac{c}{6}(1+e^{A^*}).
\end{equation}

We found that (\ref{ki}) indeed gives us the universal relation
\begin{equation}
    \sigma_1=\frac{\sigma_2}{2}+\frac{c}{6}.
\end{equation}

\section{Universal formula of EE in $d=1$}
\label{d1}
Now we move on to the configuration with $d\ge 1$ compact dimensions. To demonstrate a simpler example, we first show the $d=1$ case explicitly, since the strategy here will be essentially the same as the most general $d\ge 1$ case, and the calculational procedure is easier to understand. We set the $1+1d$ metric to be in the conformal gauge, $g_{00}=g_{11}=e^{2\phi(u_0,u_1)}$. The metric on the constant time slice is then
\begin{equation}
    ds^2=e^{2A(u_0,u_1)}\frac{dx^2}{x^2}+e^{2\phi(u_0,u_1)}(du_0^2+du_1^2),
    \label{metric1d}
\end{equation} 

where $x$ is the space coordinate on AdS$_2$ slices, $u_0$ the non-compact foliation of those AdS$_2$ slices, and $u_1$ the compact dimension $u_1\sim u_1+1$. At $u_0\rightarrow\pm\infty$ we recover AdS$_3$ times the internal compact space, which we assume it to be $S^1$. The RT surfaces are now 2d surfaces, their boundaries being 1d curves wrapping around the compact $u_1$ direction. We will see that this is drastically different from the $d=0$ case above. There are two major differences. First, in case 1 where we parametrize using $x(u_\mu)$, the conservation of the Noether currents reads $\p_\mu J_\mu=0$, which means that the currents are no longer just constants, but functions of the coordinates $u_\mu$. This yields a huge difficulty in determining $\sigma_1$. In particular, in case 1b) of $d=0$, the integration constant could be determined explicitly as $J_0=e^{A^*}$, whereas in $d\ge 1$, the conservation equation only becomes an extra constraint on the current functions. Thus we will only be able to express $\sigma_1$ in terms of an integral rather than giving an explicit value. The second difference lies in case 2 where we parametrize using $u_0(x,u_i)$, and the EOMs of the RT Lagrangian cannot be solved like in the $d=0$ case. Instead of $u_0=0$, the surface now has complicated behavior in the $(u_0,u_i)$ space, depending on the warpfactor $A(u_0,u_i)$. Once again we cannot write down the value of $\sigma_2/2$, but in the form of an integral, under the minimal surface condition.

\par
\begin{figure}[h]
    \centering
    \includegraphics[width=0.7\textwidth]{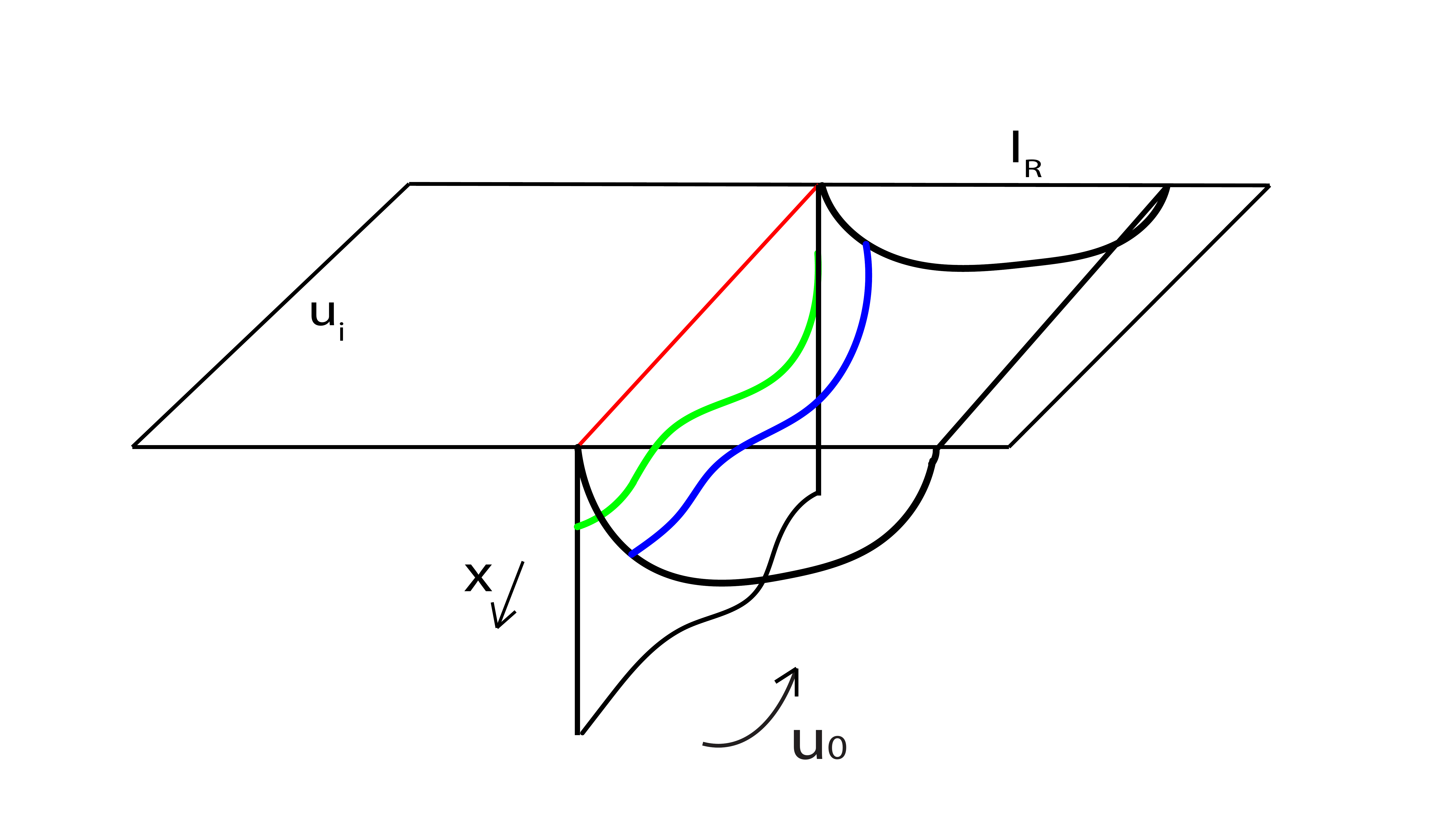}
    \caption{The curve on RT surfaces at constant $x$ slice near the interface for case 1b and case 2, with internal dimensions present. The green curve is $u_{0c}(u_1)$ on the case 2 RT surface. The blue one is $u_{0C}(u_1)$ on the case 1b RT surface. The distance from the interface is exaggerated for graphic clarity.}
    \label{fig:u0c}
\end{figure}

 We can still prove the universal relation of the $\log \epsilon$ terms in the entanglement entropy, barring those difficulties stated above. Parallel to the $d=0$ case in the previous section, we first identify how the RT surfaces in case 1b) and case 2 approaches the interface defect: $u_{0C}(u_1)=u_{0c}(u_1)$, then show that it leads to the universal relation \eqref{relation}. The way to prove \eqref{ki} is as follows: For case 2, using parametrization $u_0(u_1,x)$. The expression for $\sigma_2$ will be given by the curve $u_{0c}(u_1)$ satisfying the constraint of EOM. Then, in case 1, write down the expression of $\sigma_1$ using variation of the surface area $\mathcal{A}$ at its ends. The constraint on the boundary curve $u_{0C}(u_1)$ in case 1b is posed by $l_L=0$ (or equivalently by demanding the on-shell Lagrangian to be singular). Those two crucial curves are sketched in Figure \ref{fig:u0c}. Instead of solving for these curves explicitly, we prove that the constraints they satisfy are exactly the same. Assume that the solution to the minimal surface ODE is unique\footnote{The work on the existence of multiple solutions for minimal surface equations and its effects on the universal relation is under progress.}, then $u_{0C}(u_1)$ and $u_{0c}(u_1)$ are identified.

\subsection{Case 2}
In case 2, the region of which we calculate the EE is all of CFT$_R$, so that the 2D RT surface is attached to the CFT interface on one end and goes to infinity on the other end, as in Figure \ref{fig:u0c}. Writing down the Lagrangian in this case in terms of $du_1$ and $dx$:
\begin{equation}
    \ml=\left(\det\left(\begin{array}{cc}
            e^{2\phi}\left(\frac{\p u_0}{\p x}\right)^2+\frac{e^{2A}}{x^2}&e^{2\phi}\frac{\p u_0}{\p x}\frac{\p u_0}{\p u_1}\\e^{2\phi}\frac{\p u_0}{\p x}\frac{\p u_0}{\p u_1}&e^{2\phi}\left(\left(\frac{\p u_0}{\p u_1}\right)^2+1\right)\\
        \end{array}
    \right)\right)^{\frac{1}{2}}du_1dx
\end{equation}

and the area for the 2-dimensional RT surface will be
\begin{equation}
    \ca=\int_{0}^{1}du_1\int_0^\infty dx\  e^{\phi}\left(\frac{e^{2A}}{x^2}\left(\frac{\p u_0}{\p u_1}\right)^2+e^{2\phi}\left(\frac{\p u_0}{\p x}\right)^2+\frac{e^{2A}}{x^2}\right)^{\frac{1}{2}}.
\end{equation}

For the above area function be minimal as required for RT surfaces, notice that 
\begin{equation}
    e^{\phi}\left(\frac{e^{2A}}{x^2}\left(\frac{\p u_0}{\p u_1}\right)^2+e^{2\phi}\left(\frac{\p u_0}{\p x}\right)^2+\frac{e^{2A}}{x^2}\right)^{\frac{1}{2}}\ge e^{\phi}\left(\frac{e^{2A}}{x^2}\left(\frac{\p u_0}{\p u_1}\right)^2+\frac{e^{2A}}{x^2}\right)^{\frac{1}{2}},
    \label{inequality}
\end{equation}

and the equal sign is established if and only if $\frac{\p u_0}{\p x}=0$, i.e. the RT surface is flat on the $x$ direction (see the downward RT surface in Figure \ref{fig:u0c}). To obtain the trajectory of $u_0(u_1)$, we write down the EOM for the minimal area requirement
\begin{equation}
    \frac{\p}{\p u_1}\frac{\delta\ml}{\delta\p_{u_1}u_0}=\frac{\delta\ml}{\delta u_0}
\end{equation}

which reads
\begin{equation}
    \frac{\p (A+\phi)}{\p u_1}u_0'(1+u_0'^2)+u_0''=\frac{\p (A+\phi)}{\p u_0}(1+u_0'^2)^2.
    \label{traj}
\end{equation}

Denote this trajectory as $u_{0c}(u_1)$, so that the entanglement entropy is
\begin{equation}
    S=\frac{\ca}{4G_{4}}=\frac{1}{4G_{4}}\log L/\epsilon \int_{0}^{1}du_1e^{(A+\phi)(u_{0c}(u_1),u_1)}(1+u_{0c}'^2)^{1/2}
    \label{co2}
\end{equation}

where $L$ and $\epsilon$ are IR and UV cutoffs, and $G_{4}$ is the gravitational constant for the entire $4$-dimensional spacetime\footnote{Because the spacetime in our setup is the compact internal manifold $M^d$ non-trivially fibered over the asymptotic AdS$_3$, in general the relation between $G_{d+3}$ for the bulk spacetime and the three dimensional gravitational constant $G_3$ is not known. However, near the boundary of the bulk where spacetime factorizes to the product form AdS$_3\times M^d$, we can see $G_3$ as being Kaluza-Klein (KK) reduced from $G_{d+3}$, where they are related as $G_{d+3}=V_{M^d}\cdot G_3$ ($V_{M^d}=\int du_i\sqrt{\det g_{ij}(u_0\rightarrow\pm\infty)}$ is the volume of the compact Riemannian manifold $M^d$ near the boundary. In the case where $d=1$ and the metric is \eqref{metric1d}, we have $V_{S^1}=e^{\phi(u_0\rightarrow\pm\infty)}=1$ since $(x,t,u_0)$ forms an AdS$_3$ there). We will use this fact later along with the Brown-Henneaux relation to recover the universal relation in higher dimensions.}. This corresponds to a $-\frac{\sigma_2}{2}\log\epsilon$ term in the EE with
\begin{equation}
    \frac{\sigma_2}{2}=\frac{1}{4G_{4}} \int_{0}^{1}du_1e^{(A+\phi)(u_{0c}(u_1),u_1)}(1+u_{0c}'^2)^{1/2}.
\end{equation}

As stated above, we don't have to calculate the expression (\ref{co2}) explicitly. We just need this expression to compare with that of $\sigma_1$ and identify the relation between them.

\subsection{Generic case 1}
As mentioned in the previous section, the RT surface in case 1 is described by $x(u_0,u_1)$. The Lagrangian of it can be written as (Henceforth $\p_\mu$ will always mean $\p/\p_{u_\mu}$ for simplicity on notations, $\mu=0,1,\dots,d$)
\begin{equation}
    \begin{split}
        \ml&=\left(\mathrm{Det}\left(e^{2A}\frac{\partial_\mu x\partial_\nu x}{x^2}+e^{2\phi}\delta_{\mu\nu}\right)\right)^{1/2}\\
        &=e^\phi\left(\frac{e^{2A}}{x^2}((\partial_0x)^2+(\partial_1x)^2)+e^{2\phi}\right)^{1/2}\\
    \end{split}
    \label{lag}
\end{equation}

And the area of RT surface is
\begin{equation}
    \ca=\int_{-\infty}^\infty du_0\int_{0}^{1}du_1\ml
\end{equation}

The Lagrangian has a scale invariance under $x\rightarrow \lambda x$, which corresponds to conserved Noether current components
\begin{equation}
    \begin{split}
        J_0(u_0,u_1)&=-\frac{\delta\ml}{\delta\p_0x}\frac{dx}{d\epsilon}=-\frac{e^{2A+2\phi}}{\ml}\frac{\p_0x}{x},\\
        J_1(u_0,u_1)&=-\frac{e^{2A+2\phi}}{\ml}\frac{\p_1x}{x},\\
    \end{split}
\end{equation}

with conservation law
\begin{equation}
    \partial_0J_0+\p_1J_1=0.
\end{equation}

Written in another way, it is
\begin{equation}
    \begin{split}
        \frac{\p_0x}{x}=-\frac{e^{-A}J_0(u_0,u_1)}{\sqrt{e^{2A}-e^{-2\phi}\left(J_0^2(u_0,u_1)+J_1^2(u_0,u_1)\right)}}\\
        \frac{\p_1x}{x}=-\frac{e^{-A}J_1(u_0,u_1)}{\sqrt{e^{2A}-e^{-2\phi}\left(J_0^2(u_0,u_1)+J_1^2(u_0,u_1)\right)}},\\
    \end{split}
    \label{eom}
\end{equation}

and the on-shell Lagrangian evaluated on the solution is
\begin{equation}
    \ml=\frac{e^{A+2\phi}}{\sqrt{e^{2A}-e^{-2\phi}\left(J_0^2(u_0,u_1)+J_1^2(u_0,u_1)\right)}}.
    \label{shell}
\end{equation}

\par We always assume that the RT surface is sufficiently smooth, and $J_0,J_1$ are differentiable.

For a generic asymmetric interval, the boundary of the RT surface attaches the endpoints of the interval $[-l_L,l_R]$ with $l_L,l_R>0$. We will show that the divergent part of the entropy simply gives a $\frac{c}{6}\log l/\epsilon$ at each end of the interval, just as was found in the $d=0$ case in \cite{Karch:2021qhd}. We vary the area $\ca$ and see the $l$ dependence from the cutoff. Namely we have
\begin{equation}
    \delta\ca=\int_{0}^{1}du_1\left[\ml|_{u_0=u_c^+}\frac{\delta u_c^+}{\delta l_R}\delta l_R-\ml|_{u_0=u_c^-}\frac{\delta u_c^-}{\delta l_L}\delta l_L\right],
    \label{dca}
\end{equation}

where $u_c^\pm$ are the cutoffs at $u_0\rightarrow\pm\infty$ that we introduced at \eqref{+co}, \eqref{-co}. At $u_0=u_c^+$ where spacetime is asymptotically AdS$_3\times S^1$, we will show that $\ml$ is 1, just as in the $d=0$ case. Same argument will follow for $u_0=u_c^-$.

Near the boundary where $u_0=u_c^+$, the spacetime factors to the product of AdS$_3$ and the compact $u_1$ direction. The warpfactor becomes $e^A \sim \cosh(u_0)$ that does not depend on $u_1$, and the metric $e^\phi$ should also be independent of $u_1$, thereby a constant as $u_0\rightarrow \infty$. As mentioned above in the footnote, this is the volume of the factorized $S^1$ near the boundary. The fact that $(t,x,u_0)$ forms an AdS$_3$ at the boundary implies that $e^{\phi(u_0\rightarrow \infty)}=1$. On the other hand, the 2-dimensional RT surface in our case is a 2-dimensional minimal area submanifold embedded in AdS$_3\times S^1$, and its near-boundary behavior can be expanded in the Poincare coordinate $z$ as \cite{Robin_Graham_1999,Graham_2014}
\begin{equation}
    x(u_0,u_1)\sim x_0+x_2(u_1) z^2+x_3(u_1) z^3+...=x_0+4x_2(u_1)x_0^2e^{-2u_0}+O(e^{-3u_0})
    \label{expansion}
\end{equation}

where we have used \eqref{xz}, and $x_0=l_R$ should be the right end limit of the interval. $x_2(u_1)$ is a differentiable function on compact $u_1$ (thus being bounded), and can be determined locally by the minimal area equation \cite{Robin_Graham_1999}. The first term in the Lagrangian can be determined by the above expansion:
\begin{equation}
    e^{2A}\left(\left(\frac{\p_0 x}{x}\right)^2+\left(\frac{\p_1 x}{x}\right)^2\right)\sim 4x_0^2e^{-2u_0}\left(4x_2(u_1)^2+(x_2'(u_1))^2\right)\sim 0,
\end{equation}

where we used the asymptotic form for the warpfactor $e^A\sim e^{u_0}/2$ at $u_0\rightarrow\infty$, and the fact that $x_2$ and its derivative should be finite due to its compact support. From \eqref{lag} we get 
\begin{equation}
    \ml|_{u_0=u_c^+}=\ml|_{u_0=u_c^-}=1.
    \label{l1}
\end{equation}

Alternatively, from \eqref{expansion} the two Noether currents has behavior
\begin{equation}
    J_0,J_1\sim \frac{const.}{\ml}
    \label{jj}
\end{equation}

as $u_0\rightarrow\pm\infty$. Combining this with \eqref{shell} also gives us $\ml=1$ at the boundaries of the RT surface.

Using \eqref{+co}, \eqref{-co}, we integrate \eqref{dca} and at the cutoff Poincare coordinate $z=\epsilon$ we have \cite{Anous_2022}
\begin{equation}
    \ca=\log l_L/\epsilon+\log l_R/\epsilon+const.
    \label{315}
\end{equation}

The divergent part of the entanglement entropy is then
\begin{equation}
    S=\frac{1}{2G_4}\log l/\epsilon,
\end{equation}

where $G_4$ is again the 4-dimensional Newton's constant, and $l=l_L+l_R$ the total length of the interval. However, because here spacetime is in the product form of AdS$_3\times S^1$, the 3-dimensional gravitational constant $G_3$ can be seen as compactified from the 4D $G_4$ on the compact  $S^1$. By the standard argument of Kaluza-Klein (KK) dimensional reduction, or just by staring at the Green function for the gravitational potential, we have 
\begin{equation}
    G_4=V_{S^1}G_3=G_3,
\end{equation}

as $V_{S^1}=e^{\phi(u_0\rightarrow \infty)}=1$ near the boundary. As a result, using the BH relation \eqref{brownhenneaux} for $G_3$, the contribution to $\sigma_1$ on each end of the interval is again $\frac{c}{6}$.

\subsection{Case 1b}
The interesting case is where $l_L=0$. In this case there is a 'critical boundary' $u_{0C}(u_1)$ of the RT surface on which the denominators in the EOM are 0. As mentioned above, this boundary corresponds to the way the RT surface approaches $x=0$ with respect to different internal parameter $u_1$. One should note that this case is not just a special case of \eqref{315}. For the generic type 1 EE, the RT surfaces stretches all the way from $u_0 = \infty$ to $u_0=-\infty$; it's those two limits that drive the divergent contributions. In the special $l_L=0$ case the near-defect end of the RT surface his the boundary instead by reaching $x=0$ on the extremal $u_{0C}(u_1)$ slice, giving rise to a very different structure of UV divergent terms. See Figure \ref{fig:3rt}, \ref{fig:u0c}. More specifically, we have the minima of the functional
\begin{equation}
    D(u_{0C}(u_1),u_1)\equiv[e^{2A}-e^{-2\phi}(J_0^2+J_1^2)](u_{0C}(u_1),u_1)=0
    \label{crit}
\end{equation}

on the boundary $u_{0C}(u_1)$. That is to say, the constant $u_1$ slice of the RT minimal surface hits the interface at the AdS$_2$ slice with $u_0=u_{0C}(u_1)$.  

As the RT surface hits the boundary right on the interface at $u_{0C}(u_1)$, we still have a standard UV divergence, so a cutoff has to be introduced. Since on a constant slice we have $ds^2=e^{2A}dx^2/x^2$, at Poincare coordinate $z=\epsilon$ the cutoff reads
\begin{equation}
    e^{A_R}\epsilon=x,
\end{equation}

where $A_R=A(u_{0C}(u_1)+\delta_C,u_1)$. $\delta_C\rightarrow 0^+$ is then the cutoff on $u_0$, and we denote $u_{0C}^+=u_{0C}+\delta_C$. Since at $(u_{0C},u_1)$ the denominator $D(u_0,u_1)$ reaches its minimum 0, at the cutoff we have
\begin{equation}
    D(u_{0C}^+,u_1)=O(\delta_C^2),
\end{equation}

Now we vary the area functional and get
\begin{equation}            \delta\ca=\int_{0}^{1}du_1\left[\ml|_{u_0=u_c^+}\frac{\delta u_c^+}{\delta l_R}\delta l_R-\ml|_{u_0=u_{0C}^+}\frac{\delta (u_{0C}^+)}{\delta l_L}\delta l_L\right].
\label{da}
\end{equation}

The first term has been discussed in the previous subsection. Below we will focus on the second term. 
 The Lagrangian near the interface reads
\begin{equation}    \ml|_{u_0=u_{0C}+\delta_C}=\frac{e^{A_R+2\phi_R}}{\sqrt{D(u_{0C}^+,u_1)}}=\frac{e^{A+2\phi}(u_{0C}(u_1),u_1)}{\sqrt{D(u_{0C}^+,u_1)}}+O(\delta_C).
\end{equation}

Meanwhile, integrating the first equation of \eqref{eom} on a constant $u_1$ slice from $u_0=u_{0C}^+$, $x=e^{A_R}\epsilon$ to $u_0=u_c^+$, $x=l_R$ gives us
\begin{equation}
    \log\left(\frac{l_R}{e^{A_R}\epsilon}\right)=\int_{u_{0C}^+}^{u_c^+}du_0\frac{J_0e^{-A}}{\sqrt{D(u_0,u_1)}}.
\end{equation}

Taking an $l$ derivative on both sides then implies

\begin{equation}
    \frac{1}{l}=\frac{J_0e^{-A}}{\sqrt{D}}\bigg|_{(u_c^+,u_1)}\frac{\delta u_c^+}{\delta l}-\frac{J_0e^{-A}}{\sqrt{D}}\bigg|_{(u_{0C}^+,u_1)}\frac{\delta  u_{0C}^+}{\delta l}.
\end{equation}

The first term goes to zero by \eqref{jj}, \eqref{l1}. The second term then gives the expression of $\frac{\delta u_c^+}{\delta l}$. Omitting the $O(\delta_C)$ corrections, from \eqref{da} we get
\begin{equation}
    \frac{\delta\ca}{\delta l}=\frac{1}{l}\int_{0}^{1}du_1\frac{e^{2(A+\phi)(u_{0C}(u_1),u_1)}}{J_0(u_{0C}(u_1),u_1)}.
\end{equation}

Integrate over $l$, this near-interface side contributes
\begin{equation}
    \frac{1}{4G_4}\int_{0}^{1}du_1\frac{e^{2(A+\phi)(u_{0C}(u_1),u_1)}}{J_0(u_{0C}(u_1),u_1)}
    \label{co1b}
\end{equation}

to the divergence part of $S=A/4G_4$. 

For the $l_R$ end, as in the previous subsection, the contribution to $\mathcal{A}$ is again $\log l_R/\epsilon$.

\subsection{Relation between $\sigma_1$ and $\sigma_2/2$}

We want to identify the relation between the $\log\epsilon$ coefficients of EE of case 2 \eqref{co2} with the one of case 1b \eqref{co1b} (the part at $x=0$). Summarize them here, they are
\begin{equation}
    \frac{\sigma_2}{2}=\frac{1}{4G_4}\int_{0}^{1}du_1e^{(A+\phi)(u_{0c}(u_1),u_1)}(1+u_{0c}'^2)^{1/2},
    \label{s2}
\end{equation}

for case 2, where $u_{0c}$ satisfies the extremal trajectory condition \eqref{traj}, and
\begin{equation}
    \sigma_1=\frac{1}{4G_4}\int_{0}^{1}du_1\frac{e^{(2A+2\phi)(u_{0C}(u_1),u_1)}}{J_0(u_{0C}(u_1),u_1)}+\frac{c}{6}=\frac{1}{4G_4}\int_{0}^{1}du_1e^{A+\phi}\sqrt{1+\left(\frac{J_1}{J_0}\right)^2}\bigg|_{(u_{0C}(u_1),u_1)}+\frac{c}{6},   
\end{equation}

for case 1b, where $u_{0C}$ is solution to \eqref{crit}. We now analyze the properties of $u_{0C}$ further. Recall that under case 1b, by definition, the function $D(u_0,u_1)\equiv e^{2A}-e^{-2\phi}
(J_0^2+J_1^2)$ reaches its minimal 0 on the critical slice $u_{0C}(u_1)$. That means
\begin{equation}
    \frac{d D}{du_0}\bigg|_{u_{0C}}=\frac{dD}{du_1}\bigg|_{u_{0C}}=0
\end{equation}
 
in addition to $D|_{u_{0C}}=0$. Explicitly, this is
\begin{equation}
    \begin{split}
       \p_0(A+\phi)e^{2(A+\phi)}&=J_0\p_0J_0+J_1\p_0J_1\\ \p_1(A+\phi)e^{2(A+\phi)}&=\frac{1}{2}\left(\p_1(J_0^2+J_1^2)+u_{0C}'\p_0(J_0^2+J_1^2)\right)\\
    \end{split}
    \label{pa}
\end{equation}

on the critical slice $u_{0C}$. Meanwhile, recall that the critical boundary $u_{0C}(u_1)$ is defined in a limit fashion: it describes the behavior of the RT surface governed under \eqref{eom} when it approaches $x=0$ via constant $x$ slices. In other words, the tangent vector along a constant $x$ slice of the RT surface near the interface will give us
\begin{equation}
    dx=0=\frac{\p x}{\p u_0}du_0+\frac{\p x}{\p u_1}du_1.
\end{equation}

Taking into account the equations of motion \eqref{eom} and in the limit $x\rightarrow 0$ we have the partial derivative
\begin{equation}
    \frac{\p u_{0C}}{\p u_1}=-\frac{J_1}{J_0}.
    \label{pu0}
\end{equation}

The coefficient $\sigma_1$ is then 
\begin{equation}
    \sigma_1=\frac{1}{4G_4}\int_{0}^{1}du_1e^{A+\phi}\sqrt{1+u_{0C}'^2}\bigg|_{(u_{0C}(u_1),u_1)}+\frac{c}{6}.
    \label{s1}
\end{equation}

Now we prove the key identity $u_{0C}=u_{0c}$. This will give us the universal relation between $\sigma_1$ from case 1b and $\sigma_2$ from case 2. As mentioned above, the way we identify these two curves is to prove that $u_{0C}$ satisfies the minimal surface equation \eqref{traj} satisfied by $u_{0c}$.

Take a $u_1$ derivative on both sides of (\ref{pu0}), combined with \eqref{pa} we have
\begin{equation}
    \begin{split}
        \frac{d^2u_{0C}}{du_1^2}&=\frac{\p\frac{du_{0C}}{du_1}}{\p u_0}\frac{du_{0C}}{du_1}+\frac{\p\frac{du_{0C}}{du_1}}{\p u_1}\\
        &=\frac{-J_0\p_0J_1+J_1\p_0J_0}{J_0^2}\left(-\frac{J_1}{J_0}\right)+\frac{-J_0\p_1J_1+J_1\p_1J_0}{J_0^2}.\\
    \end{split}
    \label{20c}
\end{equation}

On the other hand, \eqref{pa} implies that
\begin{equation}
\begin{split}
    &\frac{\p (A+\phi)}{\p u_1}u_{0C}'(1+u_{0C}'^2)-\frac{\p (A+\phi)}{\p u_0}(1+u_{0C}'^2)^2\\
    &=e^{-2(A+\phi)}\bigg[\left(J_0\p_1J_0+J_1\p_1J_1-\frac{J_1}{J_0}(J_0\p_0J_0+J_1\p_0J_1)\right)\left(-\frac{J_1}{J_0}\right)\left(1+\frac{J_1^2}{J_0^2}\right)\\
    &-(J_0\p_0J_0+J_1\p_0J_1)\left(1+\frac{2J_1^2}{J_0^2}\right)\bigg]\\
\end{split}
\label{10c}
\end{equation}

Under the conservation law $\p_0J_0+\p_1J_1=0$ and \eqref{crit}, we can quickly check that \eqref{20c} and \eqref{10c} are exactly opposite. We then have
\begin{equation}
    \frac{\p (A+\phi)}{\p u_1}u_{0C}'(1+u_{0C}'^2)+u_{0C}''=\frac{\p (A+\phi)}{\p u_0}(1+u_{0C}'^2)^2.
\end{equation}

Comparing to \eqref{traj}, we see that this critical boundary $u_{0C}$ is just the trajectory $u_{0c}$ satisfying the extremal equation in case 2, barring subtleties on multiplicities of solutions that we mentioned before. Hence we identify the coefficients (\ref{s2}) under (\ref{traj}) and (\ref{s1}) under (\ref{crit}) (for the $x=0$ side contribution). In the end the universal relation  
\begin{equation}
    \sigma_1=\frac{\sigma_2}{2}+\frac{c}{6}
\end{equation}

still holds for internal compact dimension $d=1$.

\section{Higher dimensional case}
\label{dd}

Now we generalize the universal relation to the setup with arbitrary $d$ dimensional internal manifold. The most general holographic bulk has metric of the form (\ref{metric}). Below we will always abbreviate indices as $\mu=0,\dots,d$ and $i=1,\dots,d$. First we claim that there always exists a diffeomorphism $u_\mu\rightarrow \hat{u}_\mu$, such that the metric components $g_{0i}\rightarrow 0$ under this transformation. Then, we will show explicitly that the universal relation holds true in this $d$-dimensional setup with $g_{0i}=0$. The general procedure is in parallel with the previous section for $d=1$, so we won't repeat some of the subtleties and detailed arguments here.

\par First, we do the the aforementioned diffeomorphism transofrmation on our coordinates. The underlying assumption is that diffeomorphsim on the bulk spacetime manifold should not affect any physical arguments that we make. It is defined as follows \footnote{We'd like to thank Robin Graham for helpful discussions on this point.}: $\hat{u}_0=u_0$, and $\hat{u}_i$ are solutions of ODE's
\begin{equation}
    \frac{d\hat{u}_i}{du_0}=-\frac{g_{00}}{g_{0i}}(u_0,\hat{u}_i),
\end{equation}

with initial condition $\hat{u}_i(u_0=0)=u_i$, $i=1,\dots,d$. We assume that the support of $g_{0i}(u_\mu)$ is a connected region in $\mathbb{R}\times M^d$, so that we can always find solutions for the ODE where $g_{0i}$ is non-zero. This ODE will not blow up for $\hat{u}_i$, since it runs on the compact $C^\infty$-manifold, and the existence of solution relies on the non-compactness of $u_0$. The transformation of the metric is then $\hat{g}_{\mu'\nu'}=\frac{\partial u_\mu}{\p u_\mu'}\frac{\p u_\nu}{\p u_\nu'}g_{\mu\nu}$. In particular, we have
\begin{equation}
    \hat{g}_{0i}=\frac{\p u_i}{\p \hat{u}_i}g_{0i}+\frac{\p u_0}{\p \hat{u}_i}g_{00}=0.
\end{equation}

Geometrically this can be understood in the following picture. The structure we care about is the compact manifold $M^d$ with a non-compact 'time' $u_0\in \mathbb{R}$. Consider the cylinder $M^d\times \mathbb{R}$. Under the ODE's, the tangent vector $\p\hat{u}_i/\p u_0$ is orthogonal to the constant $u_0=C$ slice, which is in the $(g_{00},g_{0i})$ direction for $(u_0,u_i)$ (in other words, a gradient flow). Note that $u_0$ does not change under this transformation. In other words, for this new coordinate set $(u_0,\hat{u}_i)$ the killing vectors $\p/\p u_0$ and $\p/\p \hat{u}_i$ are orthogonal to each other, which means $\hat{g}_{0i}=0$. \\

\par Now we calculate the universal coefficient for the EE. Note that for now we keep the most general form of metric $g_{\mu\nu}$ when doing the calculation.

The metric for the spacetime is \eqref{metric}, with assumptions and conditions in section \ref{intro}. For case 2, the $d+1$-dimensional RT surface is described by $u_0(x,u_1,\dots,u_d)$. The Lagrangian (area Jacobian) of it can be written as
\begin{equation}
    \ml=\left(\det\left(\begin{array}{ccc}
            g_{00}\left(\frac{\p u_0}{\p x}\right)^2+\frac{e^{2A}}{x^2}&g_{00}\frac{\p u_0}{\p x}\frac{\p u_0}{\p u_1}+g_{01}\frac{\p u_0}{\p x}&\dots\\g_{00}\frac{\p u_0}{\p x}\frac{\p u_0}{\p u_1}+g_{01}\frac{\p u_0}{\p x}&g_{00}\left(\frac{\p u_0}{\p u_1}\right)^2+2g_{01}\frac{\p u_0}{\p u_1}+g_{11}& \\\vdots& &\ddots\\
        \end{array}
    \right)\right)^{\frac{1}{2}}dxdu_1\dots du_d
\end{equation}

After simplification and integration, the area for the RT surface becomes (remember $\p_\mu\equiv\p_{u_\mu}$)
\begin{equation}
    \ca=\int dxdu_1\dots du_d \left(\det G\left(\frac{\p u_0}{\p x}\right)^2+\frac{e^{2A}}{x^2}\sum_{\mu\nu}C_{\mu\nu}\p_\mu u_0\p_\nu u_0\right)^{1/2},
\end{equation}

where $G$ is the $(1+d)\times (1+d)$ metric matrix $G=g_{\mu\nu}(u_\lambda)$, and $C_{\mu\nu}=(-)^{\mu+\nu}M_{\mu\nu}$ are the co-factor matrices with respect to entries $\mu\nu$ of $G$. Notice that under $g_{0i}=0$,  $\det G=g_{00}C_{00}$. We assume $\det G>0$. We set the symbol $\p_0u_0=-1$, in accordance with the compact form of the equation above. Similar to the inequality \eqref{inequality}, the minimum is achieved by setting $\p_x u_0=0$, and with this we can write down the trajectory EOM for $u_i$:
\begin{equation}
    \left(\sum_i\frac{\p}{\p u_i}\frac{\delta\ml'}{\delta\p_iu_0}-\frac{\delta\ml'}{\delta u_0}\right)\bigg|_{u_{0c}}=0
\end{equation}

where
\begin{equation}
    \ml'=e^A\left(\sum_{\mu\nu}C_{\mu\nu}\p_\mu u_0\p_\nu u_0\right)^{1/2}.
\end{equation}

Explicitly, it is (denote $\p_{u_i}u_0\equiv\p_i u_0\equiv u_0^i,\p_{u_i}\p_{u_j}u_0\equiv u_0^{ij}$)
\begin{equation}
    \sum_i\p_i\left(\frac{e^A\sum_\nu C_{i\nu}u_0^\nu}{(\sum_{\mu\rho}C_{\mu\rho}u_0^\mu u_0^\rho)^{1/2}}\right)=\p_0(e^A(\sum_{\mu\rho}C_{\mu\rho}u_0^\mu u_0^\rho)^{1/2}).
    \label{dtraj}
\end{equation}

For the EE coefficient of the divergence term, we find
\begin{equation}
    \frac{\sigma_2}{2}=\frac{1}{4G_{d+3}}\int du_1\dots du_de^A\left(\sum_{\mu\nu}C_{\mu\nu}\p_\mu u_0\p_\nu u_0\right)^{1/2}\bigg|_{u_{0c}}.
    \label{ds2}
\end{equation}

Here $G_{d+3}$ is the gravitational constant in the entire $d+3$-dimensional spacetime. \\

\par For case 1, the Lagrangian for the RT surface described by $x(u_0,u_1,\dots,u_d)$ will be
\begin{equation}
    \begin{split}
      \ml=&\left(\mathrm{det}\left(e^{2A}\frac{\partial_\mu x\partial_\nu x}{x^2}+g_{\mu\nu}\right)\right)^{1/2}\\
        =&\left(\frac{e^{2A}}{x^2}\sum_{\mu\nu}C_{\mu\nu}\p_\mu x\p_\nu x+\det G\right)^{1/2}.\\
    \end{split}
    \label{lagd}
\end{equation}

The Noether current for the continuous symmetry $x\rightarrow \lambda x$ is
\begin{equation}
    J_\mu=-x\frac{\delta\ml}{\delta x}=-\frac{e^{2A}\sum_{\nu}C_{\mu\nu}\p_\nu x}{x\ml},
\end{equation}

and conservation law is $\p_\mu J_\mu=0$. After simplification and using the inverse matrix formula with cofactors, we find the EOMs
\begin{equation}
    \frac{\p_\mu x}{x}=-\frac{e^{-A}\sum_{\nu}g_{\mu\nu}J_\nu}{\sqrt{e^{2A}\det G-\sum_{\lambda\nu}g_{\lambda\nu}J_\lambda J_\nu}},
    \label{deom}
\end{equation}

and
\begin{equation}
    \ml=\frac{e^{A}\sqrt{\det G}}{\sqrt{e^{2A}-\det^{-1}G\sum_{\mu\nu}g_{\mu\nu}J_\mu J_{\nu}}}.
\end{equation}

The rest of the argument is essentially the same with the previous section. We vary the surface area $\ca=\int\ml$ with respect to $l$, and account for the contribution to $S=\frac{\ca}{4G_{d+3}}$ from each end of the RT surface. At the $u_0\rightarrow \infty$ end of the $(1+d)$D RT surface, we can write down the asymptotic expansion of $x(u_0,u_i)$ near the boundary as $u_0\rightarrow \infty$ \cite{Graham_2014}, and prove that the first term in the parenthesis in \eqref{lagd} vanishes. This gives 
\begin{equation}
    \ml|_{u_0\rightarrow \pm\infty}=\sqrt{\det G}.
\end{equation}

Furthermore, we can use KK reduction to prove $G_{d+3}=V_{M^d}G_3=G_3$. Under $g_{0i}=0$, $V_{M^d}=\sqrt{C_{00}}=\sqrt{\det G/g_{00}}$. In addition, since the spacetime is the product form AdS$_3\times M^d$ in this region, $g_{00}(u_0\rightarrow \pm\infty)=1$. By using the BH relation, each end on the ICFT away from the interface then contributes 
  \begin{equation}
      S=\frac{\ca}{4G_{d+3}}=\frac{\sqrt{\det G}}{4V_{M^d}G_3}\log l/\epsilon=\frac{c\sqrt{g_{00}(u_0\rightarrow \pm\infty)}}{6}\log l/\epsilon=\frac{c}{6}\log l/\epsilon
      \label{sss}
  \end{equation}
  
to the divergence part of EE. Again, analysis on the Noether current $J_\mu$ gives us the same result. For the special case 1b, near $x=0$, we have the 'boundary curve' of the RT surface $u_{0C}(u_i)$ as before, characterizing how each constant $u_i$ slice approaches the interface. It is described by letting the denominator of the Lagrangian be 0, i.e. 

\begin{equation}
    D(u_0,u_i)\equiv e^{2A}-\mathrm{det}^{-1}G\sum_{\mu\nu}g_{\mu\nu}J_\mu J_{\nu}=0.
    \label{dcrit}
\end{equation} 

The variation on $l$ evaluated at the cutoff $u_{0C}(u_i)+\delta_C$ gives 
\begin{equation}
    \int du_1\cdots du_d\frac{e^{2A}\det G}{{\sum_{\nu}g_{0\nu}J_\nu}}(u_{0C}(u_i),u_i)=\int du_1\cdots du_d\frac{e^{2A}\det G}{g_{00}J_0}(u_{0C}(u_i),u_i)
\end{equation}

to $\ca$. Then the coefficient in EE reads

\begin{equation}
    \sigma_1=\frac{c}{6}+\frac{1}{4G_{d+3}}\int du_1\cdots du_d\frac{e^{2A}\det G}{g_{00}J_0}(u_{0C}(u_i),u_i).
    \label{ds1}
\end{equation}\\

\par Now we identify $\sigma_1$ and $\sigma_2/2$, and more specifically, $u_{0C}$ with $u_{0c}$. First we show that if $u_{0C}=u_{0c}$ the coefficients would be identical. Along a constant $x$-slice in the region  $x\rightarrow 0$ where $u_{0c}$ and $u_{0C}$ lives, we can use EOMs (\ref{deom}) on each $2d$ slice $u_0,u_i$ with other coordinates being constant. The partial derivatives of $u_0$ on $u_i$ then have the relation
\begin{equation}
    \frac{\p u_0}{\p u_i}\equiv u_0^i=-\frac{\sum_{\lambda}g_{i\lambda}J_\lambda}{\sum_{\lambda}g_{0\lambda}J_\lambda}.
    \label{dpu}
\end{equation}

Again, set $u_0^0\equiv -1$. Plug them in (\ref{ds2}), we have
\begin{equation}
    \frac{\sigma_2}{2}=\frac{1}{4G_{d+3}}\int du_1\dots du_d\  e^{A}\left(\frac{\sum_{\mu\nu}C_{\mu\nu}\sum_{\sigma\rho}g_{\mu\sigma}J_\sigma g_{\nu\rho}J_\rho}{\sum_{\sigma\rho}g_{0\sigma}J_\sigma g_{0\rho}J_\rho}\right)^{1/2}\bigg|_{u_{0c}}.
\end{equation}

If $u_{0c}=u_{0C}$, then it satisfies the critical boundary condition (\ref{dcrit}). Under this assumption, and noticing $\sum_{\sigma}g_{\sigma\mu}C_{\sigma\nu}=\det G\delta_{\mu\nu}$, the coefficient then becomes (further simplify it by using $g_{0i}=0$)
\begin{equation}
    \begin{split}
        \frac{\sigma_2}{2}&=\frac{1}{4G_{d+3}}\int du_1\dots du_d\  \frac{e^A\sqrt{\det G}}{g_{00}J_0}\left(\sum_{\mu\nu}g_{\mu\nu}J_\mu J_\nu\right)^{1/2}\bigg|_{u_{0C}}\\
        &=\frac{1}{4G_{d+3}}\int du_1\dots du_d\  \frac{e^{2A}\det G}{g_{00}J_0}\bigg|_{u_{0C}}=\sigma_1-\frac{c}{6}.\\
    \end{split}
    \label{eq12}
\end{equation}

Now what is left is to prove that $u_{0C}$ satisfies (\ref{dtraj}), so that \eqref{ki} holds. As before, the first-order partial derivatives of $D(u_0,u_i)$ are all 0 since it reaches the minimal value 0 on the $d$-dimensional critical slice $u_{0C}$ as $x\rightarrow 0$. It implies
\begin{equation}
\begin{split}    \p_0A&=\frac{\p_0(\mathrm{det}^{-1}G\sum_{\mu\nu}g_{\mu\nu}J_\mu J_{\nu})}{2\mathrm{det}^{-1}G\sum_{\mu\nu}g_{\mu\nu}J_\mu J_{\nu}},\\
    \p_i A&=\left(\p_i+\frac{\p u_{0C}}{\p u_i}\p_0\right)\frac{(\mathrm{det}^{-1}G\sum_{\mu\nu}g_{\mu\nu}J_\mu J_{\nu})}{2\mathrm{det}^{-1}G\sum_{\mu\nu}g_{\mu\nu}J_\mu J_{\nu}}.\\
\end{split}
\label{dpa}       
\end{equation}

The goal is to plug $u_{0C}$ into the PDE (\ref{dtraj}) for $u_{0c}$ and use identities \eqref{dcrit}, (\ref{dpa}), (\ref{dpu}) to prove it holds true. After explicitly expanding (\ref{dtraj}) it becomes
\begin{equation}
    \begin{split}
       \sum_{\nu\mu\rho}u_0^\mu u_0^\rho\bigg[&\sum_\lambda C_{\lambda\nu}C_{\mu\rho}u_0^\nu\p_\lambda A-2C_{0\nu}C_{\mu\rho}u_0^\nu\p_0A+\sum_\lambda(\p_\lambda C_{\lambda\nu}C_{\mu\rho}-1/2C_{\lambda\nu}\p_\lambda C_{\mu\rho})u_0^\nu\\
       &+(C_{0\nu}\p_0C_{\mu\rho}-2\p_0C_{0\nu}C_{\mu\rho})u_0^\nu+\sum_\lambda(C_{\lambda\nu}C_{\mu\rho}-C_{\lambda\mu}C_{\nu\rho})u_0^{\lambda\nu}\bigg]=0\\
    \end{split}
    \label{expand}
\end{equation}

where the second order derivatives are
\begin{equation}
    u_0^{\mu\nu}=-(u_0^\nu\p_0+\p_\nu)\frac{\sum_\theta g_{\mu\theta} J_\theta}{g_{00}J_0}.
\end{equation}

The partial derivatives of the determinant identity $\delta_{\nu\lambda}\p_\sigma\det G=\sum_\mu \p_\sigma(g_{\mu\nu}C_{\mu\lambda})$ is again utilized. We can verify using Mathematica that (\ref{expand}) holds true under the assumptions $g_{0i}=0,i=1,\dots,d$. Utilizing the aforementioned diffeomorphism, we can put our metric in this form, without changing any physics. Thus the relation between the two coefficients still holds in the most general holographic setup with the internal manifold of any dimension.

\section{Half-BPS Six Dimensional (0,4) Supergravity Solution}
\label{ex}
In this section we demonstrate an analytic example of a holographic dual setup consisting of asymptotic AdS$_3$ times a compact manifold of $d=3$. Specifically, it is the half-BPS solution in six dimensional (0,4) supergravity with $SO(2,1)\times SO(3)$ symmetry and $m$ tensor multiplets \cite{Chiodaroli_2011}, whose spacetime manifold is $AdS_2\times S^2$ warped over a Riemann surface $\Sigma$ with boundary. This is the generalization of the dilatonic  Janus solution with a 3 dimensional metric \cite{Bak:2007jm,Karch:2021qhd}, or in our notation $d=0$.

The spacetime manifold has two regions where it asymptotically becomes $AdS_3\times S^3$, as per the requirement for the general holographic setup at $u_0\rightarrow\pm\infty$. The non-compact coordinate lies in the Riemann surface $\Sigma$, which in this example we demand to be the upper half plane for simplicity, and is parametrized by the complex coordinate $w$. In this case where $\Sigma$ has a boundary, the aforementioned asymptotic regions occur at boundary points $x_1,x_2\in \mathbb{R}$. In the following we will sum up the analytical results of the ansatz for the metric of this six dimensional solution \cite{Chiodaroli_2011}.

\par In the orthogonal frame, the metric is
\begin{equation}
    ds^2=f_1^2ds_{AdS_2}^2+f_2^2ds_{S^2}^2+\rho^2ds_{\Sigma}^2,
\end{equation}

where $f_1,f_2,\rho$ are real functions depending only on the coordinates of $\Sigma$. It automatically has the form $g_{0i}=0,i=0,1,2,3$. The radii of both asymptotic AdS$_3$ and $S^2$ are chosen to be unity. $f_1$ plays the role of a warpfactor. We use an auxiliary real harmonic function $H(w)$ and a $SO(2,m)$ vector of locally holomorphic functions $\lambda^A(w),A=1,\dots,m+2$ to characterize $f_1,f_2,\rho$ as follows:
\begin{equation}
    \begin{split}
        H&=\sum_{n=1}^{2}\frac{ic_n}{w-x_n}+c.c.,\\   \lambda^A&=\frac{1}{\p_wH}\sum_{n=1}^{2}\left(\frac{-i\kappa_n^A}{(w-x_n)^2}+\frac{-i\mu_n^A}{w-x_n}\right),\\
    \end{split}
    \label{hl}
\end{equation}

and
\begin{equation}
    \begin{split}
        f_1^4&=H^2\frac{\lambda\cdot\bar{\lambda}+2}{\lambda\cdot\bar{\lambda}-2},\\
        f_2^4&=H^2\frac{\lambda\cdot\bar{\lambda}-2}{\lambda\cdot\bar{\lambda}+2},\\
        \rho^4&=\frac{|\p_wH|^4}{16H^2}(\lambda\cdot\bar{\lambda}-2)(\lambda\cdot\bar{\lambda}+2),\\
    \end{split}
    \label{fr}
\end{equation}

where $c_1,c_2>0$ and $\kappa_n^A,\mu_n^A$ are real numbers. $m\ge 0$ is a integer. They satisfy the relation (Note that $\kappa_1, \kappa_2,\mu_1,\mu_2$ are $(m+2)$-vectors):
\begin{equation}
    \begin{split}
        \kappa_1\cdot\kappa_1&=2c_1^2\\
        \kappa_2\cdot\kappa_2&=2c_2^2\\
        \kappa_1\cdot\mu_1&=0\\
        \kappa_2\cdot\mu_2&=0\\
        \mu_1^2=\mu_2^2&=4c_1c_2-2\kappa_1\cdot\kappa_2.\\
    \end{split}
    \label{km}
\end{equation}

First we rephrase the metric in terms of the more familiar form parametrized by non-compact $u_0$ and compact $u_1,u_2,u_3$ coordinates. Set $w=e^{u_0+iu_1}$, and $u_2,u_3$ are the standard spherical coordinates for the $S^2$. Note that under this set of coordinates, we only consider $0<u_2<\pi$, in other words, this atlas of the internal compact manifold excludes the north and south poles. Of course we can always change a set of atlas with smooth transition function. Again, the asymptotic AdS$_3$ region for our previous coordinates in (\ref{metric}) was $u_0\rightarrow\pm\infty$, while in the present coordinates the two AdS$_3$ regions are at $(u_0,u_1)=(\log |x_i|,\pi^{\mathrm{sgn}(x_i)})$. The metric then reads
\begin{equation}
    ds^2=f_1^2\frac{dx^2-dt^2}{x^2}+\rho^2e^{2u_0}(du_0^2+du_1^2)+f_2^2(du_2^2+\sin^2 u_2du_3^2).
\end{equation}

We can see that it is a diagonal metric. Nevertheless each of the diagonal components depends on $u_0$ and some of the compact coordinates, so the bulk spacetime is not simply the product form with asymptotic AdS$_3\times M^d$, except near the boundary. The co-factors are ($C_\mu\equiv C_{\mu\mu}$ since non-diagonal ones vanish)
\begin{equation}
    \begin{split}
        C_0=f_1^2f_2^4\rho^2e^{2u_0}\sin^2u_2&,\   C_1=f_1^2f_2^4\rho^2e^{2u_0}\sin^2u_2\\
        C_2=f_1^2f_2^2\rho^4e^{4u_0}\sin^2u_2&,\   C_3=f_1^2f_2^2\rho^4e^{4u_0}.\\
    \end{split}
\end{equation}

For case 2 with 4-dimensional RT surface $u_0(x,u_1,u_2,u_3)$, according to \eqref{ds2} the coefficient of EE is 
\begin{equation}
\begin{split}
    \frac{\sigma_2}{2}=\frac{1}{4G_6}\int_0^\pi du_1\int_0^\pi du_2\int_0^{2\pi}du_3 &|f_1^2f_2\rho e^{u_0}|\big(f_2^2\sin^2u_2(1+(\p_1u_0)^2)\\
    &+\rho^2e^{2u_0}\sin^2u_2(\p_2u_0)^2+\rho^2e^{2u_0}(\p_3u_0)^2\big)^{\frac{1}{2}}\bigg|_{u_{0c}}.\\
\end{split}
\label{6ds2}
\end{equation}

{The minimal boundary curve $u_{0c}$ satisfies}
\begin{equation}
    \sum_{i=1}^3\p_i\left(\frac{|f_1| C_{i}u_0^i}{\left(\sum_{\mu=0}^4 C_{\mu}(u_0^\mu)^2\right)^{1/2}}\right)=\p_0\left(|f_1|(\sum_{\mu}C_{\mu}(u_0^\mu)^2)^{1/2}\right).
    \label{traj6d}
\end{equation}

For case 1 with the $(1+d)$ dimensional RT surface $x(u_\mu)$, we can write out the Lagrangian for it as
\begin{equation}
    \ml=\rho^2f_2^2e^{2u_0}|\sin u_2|\left(\frac{f_1^2}{x^2}\left(\frac{(\p_0x)^2+(\p_1x)^2}{\rho^2e^{2u_0}}+\frac{(\p_2x)^2}{f_2^2}+\frac{(\p_3x)^2}{f_2^2\sin^2u_2}\right)+1\right)^{\frac{1}{2}},
\end{equation}

and the Noether currents $J_\mu=-\frac{C_{\mu}\p_0 x}{x\ml}$ for $\mu=0,1,2,3$. The contribution at each end near the boundary is $c/6$, given by the factorization of spacetime at the boundary, KK reduction for $G_6$, and the BH relation for $G_3$, as in the previous section. We can verify the fact that $f_1(u_0\rightarrow\pm\infty)\sim e^{\pm u_0}/2$, and $\rho^2e^{2u_0}(u_0\rightarrow\pm\infty)\sim 1$. According to (\ref{ds1}), the case 1b) coefficient is then
\begin{equation}
    \sigma_1=\frac{c}{6}+\frac{1}{4G_6}\int_0^\pi du_1\int_0^\pi du_2\int_0^{2\pi}du_3\frac{f_1^2f_2^4\rho^2e^{2u_0}\sin^2u_2}{J_0}(u_{0C})
    \label{6ds1}
\end{equation}

on the critical boundary $u_{0C}$ satisfying
\begin{equation}
    f_1^2f_2^4\rho^4e^{4u_0}\sin^2u_2-\rho^2e^{2u_0}(J_0^2+J_1^2)-f_2^2J_2^2-f_2^2\sin^2u_2J_3^2=0.
    \label{crit6d}
\end{equation}

Following the general procedure, notice that under \eqref{crit6d} the case 2 coefficient \eqref{6ds2} simplifies to exactly the second term of \eqref{6ds1}. The rest just amounts to show the equivalency of (\ref{traj6d}) with (\ref{crit6d}). The partial derivatives $\p_{u_i}u_0$ near the interface in terms of the Noether currents are
\begin{equation}
    u_0^1=-\frac{J_1}{J_0},u_0^2=-\frac{f_2^2J_2}{\rho^2e^{2u_0}J_0},u_0^3=-\frac{f_2^2\sin^2u_2J_2}{\rho^2e^{2u_0}J_0}.
    \label{u0i}
\end{equation}

Indeed, the $u_{0c}$ minimal equation simplifies under \eqref{u0i}, \eqref{crit6d} and its derivatives to
\begin{equation}
    \begin{split}
        \sum_\mu (u_0^\mu)^2\bigg[&\sum_{\nu}C_\mu\nu u_0^\nu \frac{\p_\nu |f_1|}{|f_1|}-2C_0C_\mu\frac{\p_0 |f_1|}{|f_1|}\\
        +&\sum_{\nu}(\p_\nu C_\nu C_\mu-1/2C_\nu\p_\nu C_\mu)u_0^\nu+(C_0\p_0C_\mu-2\p_0C_0C_\mu)\bigg]=0.\\
    \end{split}
    \label{final}
\end{equation}

The simplest $\kappa$ and $\mu$ in \eqref{km} are for $m=0$ and $\kappa_i=(\sqrt{2}c_i,0), \mu_i=(0,0), i=1,2$. Plugging into the parameters \eqref{hl}, \eqref{fr}, the above equation \eqref{final} holds true trivially. For general integer $m$ and $m+2$-vectors $\kappa$, $\mu$ satisfying \eqref{km}, we have used Mathematica to verify that \eqref{final} is correct at least for $m=0,1,2$. We can safely say that in this way we prove the universal relation 
\begin{equation}
    \sigma_1=\frac{\sigma_2}{2}+\frac{c}{6}
\end{equation}

in this holographic setup with $d=3$.

\section{Unequal Central Charges}
\label{uq}
The central charges of the left and right CFTs are in general not the same, as the symmetry of an ICFT allows $c$ to jump across the interface. The central charge entered our calculation via the BH relation \eqref{brownhenneaux} with the radius AdS$_3$ curvature $L$ restored
\begin{equation}
    \frac{G_3}{L}=\frac{3}{2c}.
    \label{newbh}
\end{equation}

So in the holographic picture unequal central charges actually correspond to two different radii $L_L$ and $L_R$ of the AdS$_3$ on the left and right. The two near-boundary limits $u_0\rightarrow\pm\infty$ are now asymmetrical, since the product form of spacetime on the left/right side is AdS$_3^{L_L/L_R}\times M^d$. The warpfactor now becomes \cite{Karch:2021qhd} $e^A=L\cosh \frac{u_0}{L}$ for an AdS$_3$ with radius $L$, but the results will be parallel to when we set all the $L=1$, since the length units in $e^A$ only appears explicitly when we use the cutoff procedure. Hence it is justified to only use the different curvature radii for the left- and right-hand sides in the final step.

The only subtlety is in generic case 1 RT surfaces. But we will see that the effect of different AdS$_3$ radius $L$ is exactly canceled out. When calculating the near-boundary contribution $\frac{c}{6}$ to the EE away from the interface, we used the fact that since in this region we have pure AdS$_3$, the metric component $g_{00}$ of $u_0$ becomes 1. More generally, for a region of pure AdS$_3$ with radius $L$, we should have $g_{00}=1/L^2$ in that region. For simplicity we assume it is the right side near-boundary region $u_0\rightarrow\infty$. Following the notations in section \ref{dd} and \eqref{sss}, the coefficient of the divergence part of EE is then
\begin{equation}
    \frac{1}{4G_{d+3}}\sqrt{\det G}=\frac{1}{4V_{M^d}G_3}\sqrt{\det G}=\frac{1}{4G_3}\sqrt{g_{00}}=\frac{c}{6L}\cdot L=\frac{c_R}{6}
\end{equation}

under the KK reduction on volume $V_{M^d}=\sqrt{\det G/g_{00}}$. The left side follows the same argument. 

Case 2 doesn't change at all, while in case 1b we have to specify which side we choose for the one-sided interval. Without loss of generality we will choose the right side. The second term in \eqref{ds1} will not change. We can see that the universal relation still holds as
\begin{equation}
    \sigma_1=\frac{\sigma_2}{2}+\frac{c_R}{6}.
\end{equation}

Only this time the $c$ is replaced by $c_R$, since we specified the interval to be in CFT$_R$ and ends on the interface. Notice here the integral in (\ref{ds1}) and (\ref{ds2}) both have units of length. After dividing by the Newton constant $G_{d+3}$, they yield the dimensionless entanglement entropy.

\section{Further outlook}
\label{ol}
We have shown that the universal relation between the case 1b and case 2 EE coefficients holds true for additional compact dimensions aside from the noncompact AdS$_3$, using the holographic picture. We still wish to complete the proof of this conjecture, using CFT tools instead of calculating RT surfaces. Previously some work has been done on this subject for ICFT \cite{Sakai:2008tt}, or for BCFT if we see it as a folded ICFT \cite{Cardy:2004hm,Calabrese:2009qy}. The replica trick was used when dealing with field theory derivation of the EE. 

For calculation of the case 2 EE, we can write it down analytically for simple CFT with conformal interface. The setup is sketched briefly as follows \cite{Sakai:2008tt,brehm2015entanglement}:
\par For the replicated $K$-Riemann surface $\mathcal{R}_K$ with twist fields, the complex coordinate is $w$, and branch cut is $w=0$ goes to $\infty$. Now we go to coordinate $z=\log w$, and denote the UV and IR cutoff $|w|=\epsilon, L$. Then $\mathcal{R}_K$ is mapped to a rectangle $(\log L/\epsilon, 2\pi K)$, with interfaces sitting at $2K$ lines $\mathrm{Im} z=(2m-1)\pi/2,m=1,...,2K$. Applying the periodic boundary condition on the Re$z$ direction, and setting $q=e^{-2\pi t}$ with $t=\pi/2|\log L|$, one can write done the path integral explicitly as
\begin{equation}
    Z(K)=\mathrm{Tr}_1\left(I_{12}q^{L_0^2+\tilde{L}_0^2}(I_{12})^\dagger q^{L_0^1+\tilde{L}_0^1}I_{12}\cdots (I_{12})^\dagger q^{L_0^1+\tilde{L}_0^1}\right).
    \label{pi}
\end{equation}

The EE for one of the CFT's (case 2) is then
\begin{equation}
    S=(1-\partial_K)\log Z(K)\big|_{K\rightarrow 1}.
\end{equation}

The dynamics of the interface depends crucially on the energy transmission coefficient $\mathcal{T}$ of the interface \cite{Meineri_2020}, and so does the Virasoro generators. The folding trick is often used to simplify the problem especially when symmetric interval is involved, turning the interface to a conformal boundary which is the tensor state of the two CFT's. In some special cases, such as the $c=1$ compact boson, or its supersymmetric version, we can write down the explicit expressions for the interface operator $I_{12}$ from the boundary tensor product theory of two free bosons. Then the path integral can be evaluated, and so is the EE. In the end we will find that the physical quantity determining the universal coefficients of the $\log\epsilon$ terms in the EE is also the transmission coefficients $\mathcal{T}$ \cite{Gutperle_2016}. Summarizing, we can write down analytically the universal coefficients $\sigma_2(\mathcal{T})$ for simple tractable CFT's, and using the folding trick, $\sigma(\mathcal{T})$ for a symmetric interval in case 1.

Some current work currently under progress are listed below:
\begin{itemize}
\item In a more general interface setting, the expression of $I_{12}$ is lacking, and the partition function \eqref{pi} is yet to be spelled out explicitly. Even for the simple $c=1$ compact boson with permeable interface, the physical meaning of doing the folding trick is in question in the first place if we are calculating the EE of an asymmetric interval (generic case 1). Nevertheless, EE of generic case 1 intervals will be an interesting subject to study using CFT techniques such as the replica trick.

\item An obvious further generalization of our work is to investigate the universal relation of holographic EE of a general $(n+1)$ dimensional ICFT. The intervals would be $n$ dimensional spacelike regions, and one would believe similiar relations hold for the universal $\log l/\epsilon$ terms  in EE of certain co-dimensional 1 subregions.
\end{itemize}

\section*{Acknowledgments}

Special thanks to Robin Graham for very helpful discussions. The work of AK was supported, in part, by the U.S.~Department of Energy under Grant DE-SC0022021 and by a grant from the Simons Foundation (Grant 651678, AK).

\bibliographystyle{JHEP}
\bibliography{2djhep}

\end{document}